\documentclass[11pt]{article}
\usepackage{graphicx}
\usepackage[margin=1.25in]{geometry}
\usepackage[usenames,dvipsnames]{color}
\usepackage{url}
\usepackage[colorlinks = true,
            linkcolor = blue,
            urlcolor  = blue,
            citecolor = blue,
            anchorcolor = blue]{hyperref}

\usepackage{todonotes}
\usepackage{lineno}

\makeatletter
\renewcommand\paragraph{\@startsection{paragraph}{4}{\z@}%
            {-2.5ex\@plus -1ex \@minus -.25ex}%
            {1.25ex \@plus .25ex}%
            {\normalfont\normalsize\bfseries}}
\makeatother

\newcommand\pubnumber{Preprint}
\newcommand\pubdate{\today}

\def\Title#1{\begin{center} {\LARGE #1 } \end{center}}
\def\Author#1{\begin{center}{ \sc #1} \end{center}}
\def\Address#1{\begin{center}{ \it #1} \end{center}}

\newcommand\pubblock{\rightline{\begin{tabular}{l} \pubnumber\\
         \pubdate \end{tabular}}}
\newenvironment{Abstract}{\begin{quotation} \begin{center}
                       ABSTRACT
     \end{center}\bigskip  }{\end{quotation}}

\newcommand{\cefgroup}{6}
\usepackage{dashrule}

\newenvironment{recs}
    {
   \noindent \begin{minipage}{\textwidth} \rule{\textwidth}{2mm}}
  {\rule{\textwidth}{2mm}  \end{minipage}}

\newcommand{\rec}[3]{
\nobreak \noindent \textbf{CEF0\cefgroup~Recommendation #1 -- #2}\\ 
\nobreak \hdashrule{\textwidth}{0.5pt}{2pt}
\nobreak #3\\
}

\newcommand{\subrec}[2]{
\nobreak \noindent \textbf{CEF0\cefgroup~Recommendation #1 -- #2}\\
}

\newcommand\snowmass{\begin{center}\rule[-0.2in]{\hsize}{0.01in}\\\rule{\hsize}{0.01in}\\
\vskip 0.1in Submitted to the  Proceedings of the US Community Study\\ 
on the Future of Particle Physics (Snowmass 2021)\\ 
\rule{\hsize}{0.01in}\\\rule[+0.2in]{\hsize}{0.01in} \end{center}}

\begin{document}

\pubblock

\Title{Summary Report\\Public Policy \& Government Engagement\\
Community Engagement Frontier Topical Group 06\\
Snowmass 2021\\
}

\bigskip 

\Author{Rob Fine$^1$, Louise Suter$^2$}
\Address{$^1$Los Alamos National Laboratory}
\Address{$^2$Fermi National Accelerator Laboratory}

\medskip

\begin{Abstract}
This document is the Snowmass summary report for the Public Policy \& Government Engagement topical group within the Community Engagement Frontier. This report discusses how the U.S. High Energy Physics (HEP) community currently engages with government at all levels and provides recommendations for how the execution of these activities can be improved and for how the scope of existing activities can be expanded. This includes the current HEP Congressional government engagement ``DC trip'', materials produced for communication with government officials, government engagement in areas other than HEP funding, and interactions with the funding agencies, Executive Office of the President, and state and local governments.
\end{Abstract}

\snowmass
 
\clearpage

\section{Introduction}
\label{sec:intro}

This document has been prepared as the summary report of the Snowmass group focused on Public Policy \& Government Engagement, Community Engagement Frontier topical group \#6 (CEF06). The charge of CEF06 was to review how the High Energy Physics (HEP) community engages with government at all levels and all related topics. Among the topics investigated were how public policy impacts members of the HEP community and the HEP community at large and the issue of awareness within the HEP community of direct community-driven engagement of the U.S. federal government. We note that in the previous Snowmass process, the topics which now fall under the purview of CEF06 were then included in the broader ``Community Engagement and Outreach'' group whose work culminated in the recommendations outlined in Ref. \cite{snowmass13recs}.

The organization of this report is as follows. The remainder of this section provides a high-level summary of the proceedings of CEF06, provides context for this report concerning the CEF06 contributed papers, defines key terminology, and lists the high-level recommendations detailed throughout the remainder of the report. Section \ref{sec:funding_summary} provides an overview of how HEP is funded in the U.S. and the role that community-driven government engagement plays in that process. Section \ref{sec:future_group} discusses who is accountable to take action of these recommendations and \ref{sec:community_unity} describes the importance of unity of messaging while engaging in these efforts. Section \ref{sec:current_advocacy} describes existing activities focused on engagement of the U.S. federal legislature on the topic of funding for HEP and recommendations for improving the efficacy of these activities. Section \ref{sec:outreach_materials} describes the content of the HEP community communication materials used in HEP government engagement activities and the effort that produces these materials. Section \ref{sec:non_funding_topics} describes the state of community-driven engagement on non-budgetary topics and the potential for utilizing external resources to expand activities in this area. Section \ref{sec:funding_agencies} describes the state of relations between the HEP community and funding agencies. Finally, Section \ref{sec:expanded_engagement} describes ideas for how HEP community-driven government engagement can expand beyond activities targeting the U.S. federal legislature.

\subsection{History of CEF06 Activities}

Throughout 2020, CEF06 held bi-weekly ``Town Hall'' meetings and discussion-focused ``Community Chats''. The records of these activities can be accessed on Indico \cite{cef06_indico}. Following the 2021 pause in Snowmass proceedings, weekly meetings within CEF06 focused on organizing the feedback received during 2020 into detailed summary documents. This effort formed the basis for the contributed papers prepared by the group, which are described below.

\subsection{Contributed Papers}

The culmination of CEF06 activities during the 2021-2022 Snowmass process was the development by the group of three ``contributed papers''. These papers serve as the record of the proceedings of CEF06 and provide further details in many areas than this summary report includes. The three reports are:

\begin{itemize}
    \item Snowmass 2021 Community Engagement Frontier 6: Public Policy and Government Engagement Congressional Advocacy for HEP Funding (The “DC Trip”) \cite{cef06paper1},
    \item Snowmass 2021 Community Engagement Frontier 6: Public Policy and Government Engagement Congressional Advocacy for Areas Beyond HEP funding \cite{cef06paper2}, and
    \item Snowmass 2021 Community Engagement Frontier 6: Public Policy and Government Engagement:  Non-congressional government engagement \cite{cef06paper3}.
\end{itemize}

\subsection{Definitions}

Throughout this report we utilize the following terms, which we define here for the convenience of the reader:

\begin{itemize}
    \item \textbf{AAAS} - American Association for the Advancement of Science 
    \item \textbf{AIP} - American Institute for Physics 
    \item \textbf{APS} - American Physical Society
    \item \textbf{Administration} - When capitalized, referring to a general or specific Presidential Administration; the staffing of the executive branch of the federal government
    \item \textbf{``Ask''} - An appropriations request made by the HEP community for particular funding levels for DOE OS and NSF
    \item \textbf{CEF06} - Community Engagement Frontier topical group \#6; the Snowmass group on Public Policy \& Government Engagement
    \item \textbf{``DC Trip''} - The annual HEP community advocacy trip to Washington, D.C.
    \item \textbf{DOE} - Department of Energy
    \item \textbf{EOP} - Executive Office of the President
    \item \textbf{HEP} - High Energy Physics
    \item \textbf{HEPAP} - High Energy Physics Advisory Panel; convened by DOE and NSF to provide advise on HEP priorities
    \item \textbf{NSF} - National Science Foundation
    \item \textbf{OMB} - Office of Management and Budget (within EOP) 
    \item \textbf{OS} - (DOE) Office of Science 
    \item \textbf{OSTP} - Office of Science and Technology Policy (within EOP)
    \item \textbf{OHEP} - (DOE OS) Office of High Energy Physics
    \item \textbf{PBR} - The President's Budget Request; the federal budget proposed by the President which Congress considers while forming its own budget
    \item \textbf{P5} - The Particle Physics Project Prioritization Panel and the report they produce
    \item \textbf{SLUO} - SLAC Users Organization
    \item \textbf{UEC} - Fermilab Users Executive Committee
    \item \textbf{USLUA} - U.S. Large Hadron Collider (LHC) Users Association 
    \item \textbf{WHIPS} - The Washington-HEP Integrated Planning System; framework used to coordinate logistics of HEP community advocacy
\end{itemize}



\subsection{Recommendations}

The activities of CEF06 have culminated in a number of recommendations for how the HEP community should engage with the government over the next decade. High-level recommendations are highlighted below, and are reproduced throughout this report where they are contextualized and, in cases, elaborated on through sub-recommendations (Recommendations 2, 4, and 7).

\vspace{2ex}

\begin{recs}
    \rec{1}{Representatives of APS DPF, HEPAP, and the user groups, as appropriate, should have dedicated discussions to determine what actions can be taken to advance the recommendations outlined in this report.}{There is limited bandwidth in the current infrastructure for investing resources into additional government engagement activities, and the recommendations here cover a broader scope than existing community-driven advocacy. These discussions should specifically include considering the option of forming an HEP community government engagement group, composed of elected community representatives and policy experts, that is explicitly responsible for developing expanded government engagement capabilities.}
%
    \rule{\textwidth}{0.5mm}
    \rec{2}{HEPAP must build community unity around the 2023 P5 plan and develop a clear messaging strategy spanning the next 10 years. 
}{Impactful community-driven advocacy and strong support for HEP within Congress are dependent on a unified community message that reflects the priorities outlined in the P5 report. Dedicated efforts to engage the HEP community and all relevant stakeholders are required to build and maintain this unity as the 2023 P5 report is developed, rolled-out, and implemented.}

\end{recs}
\newpage

\begin{recs}
    \rec{3}{Representatives of UEC, SLUO, USLUA, and APS DPF should facilitate discussions to consider the formation of a more formal ``HEP Congressional advocacy'' group to assume responsibility for organizing the annual advocacy trip to Washington, D.C.}{The formalization of such a group has the potential to streamline the adoption of other recommendations in this report and to elevate the profile of this important effort within the HEP community. Any discussions on this topic should include other leaders of the HEP community and provide mechanisms for feedback from community members.}
    \rule{\textwidth}{0.5mm}
    \rec{4}{The HEP Congressional advocacy group must continue to support, and should aim to grow, the annual HEP community-driven advocacy activities.}{The annual community-driven advocacy effort is essential to increase knowledge and interest of HEP in Congress. Participation in these activities should be encouraged within the community, and community leaders should support efforts for continued development and growth of this effort.}
    \rule{\textwidth}{0.5mm}
    \rec{5}{The HEP Congressional advocacy group must continue support for the development and maintenance of HEP Communication materials, working with the 2023 P5 chair and other experts as needed.}{High quality and well-developed communication and outreach materials are essential for effective government outreach, and their quality reflects directly on how our field is perceived. Investment in these important resources must be continued, and they should be made more broadly available within the community.}
    \rule{\textwidth}{0.5mm}
    \rec{6}{The HEP Congressional advocacy group and APS DPF executive committee should identify an existing community group or create a new one to take ownership of strengthening connections between the HEP community and science and physics societies, including APS, AIP, and AAAS, and continuing conversations about how the HEP community can engage in advocacy for topics beyond HEP funding.}{The purpose of strengthening these connections should be to leverage external expertise on advocating for DEI, immigration policy, R\&D, basic science funding reform, and other areas that impact HEP. Maintaining strong connections between the HEP community and these external groups will expand the resources easily available to HEP community members in these areas. The HEP group tasked with strengthening these connections to external groups should also facilitate an expanded discussion around the potential for HEP-specific community advocacy in these areas.}
\end{recs}
\newpage

\begin{recs}
    \rec{7}{DOE and NSF should improve existing communication channels and create new ones, as necessary, to enable improved HEP community feedback to the funding agencies.}{HEP community members have substantive opinions about how our community interfaces with the funding agencies and lack sufficient avenues to provide feedback to them. Any steps taken to improve communication should include the development of mechanisms for community-level, individual, and anonymous feedback to be provided on topics including the climate of the field and the granting process. Transparency about what feedback is provided and what actions are taken to address feedback should be a key element of the solution.}
    \rule{\textwidth}{0.5mm}
    \rec{8}{The HEP Congressional advocacy group should work to improve HEP community engagement of the executive branch, especially OMB and OSTP.}{Steps should be taken to optimize community-driven advocacy targeted at these groups and to separate the logistics of this effort from the community-driven advocacy targeted at Congress. Meetings with these agencies should be timed for maximum impact on the development of the President's Budget Request. Specialized outreach materials and messaging points should be developed, and HEP community members meeting with these agencies should receive specialized training.}
    \rule{\textwidth}{0.5mm}
    \rec{9}{The HEP Congressional advocacy group and APS DPF executive committee should facilitate discussions to explore the potential advantages to systematic engagement of local and state governments.}{The goal of these discussions should be understanding the potential costs and benefits of allocating resources for such activities. Specific locations of particular relevance to HEP activities should be identified to guide these discussions.}
\end{recs}
\section{HEP Funding and HEP Community-Driven Advocacy}
\label{sec:funding_summary}

\subsection{How HEP is Funded in the United States}
\label{sec:hep_funding}

To constructively discuss the role of community advocacy in funding High Energy Physics (HEP) in the United States, it is essential to briefly discuss the U.S. federal budget process. The Department of Energy (DOE), through its Office of Science (OS), and the National Science Foundation (NSF) fund HEP. Funding for these executive branch agencies is provided for in the U.S. federal budget on an annual basis. The construction of the federal budget is a lengthy process that can be summarized by roughly breaking it down into three key steps. 

\begin{enumerate}
    \item \textbf{The President proposes a budget.} The President's Budget Request (PBR) is based on input from executive branch agencies and is coordinated by the Office of Management and Budget (OMB).
    \item \textbf{Congress legislates a budget.} Both entities within the legislative branch (the House of Representatives and the Senate) propose budgets, taking into account their priorities and the priorities of the Administration (reflected in the PBR). The House and Senate budgets are then reconciled into a single Congressional budget which forms the basis for authorizations and appropriations to fund executive branch agencies. The authorization step dictates how each agency may utilize its funding, with variable degrees of specificity, while the appropriations bill allocates the funding.
    \item \textbf{OMB appropriates funds to the executive branch agencies.} This process is dictated by Congress's authorizations and appropriations bills to carry out specific programs, projects, and activities in line with agency and Administration priorities.
\end{enumerate}

The President's budget request is formulated using policy guidance from OMB informed by the previous year's budget. Based on this guidance and community input, each agency (DOE and NSF in the case of HEP) submits proposals to OMB. OMB revises and synthesizes these proposals, accounting for Administration priorities, which the President then reviews and transmits to Congress. After OMB has submitted the proposal to the President, agencies have a period during which they may appeal revisions made by OMB. The Congressional budget uses the President's budget request as input, but the appropriated budget often differs from that initial request. 

Historically HEP community-driven advocacy efforts have focused on step (2), \textit{i.e.}, advocacy aimed at Congress. Working with policy experts, the HEP community formulates a Congressional appropriations request for DOE OS HEP and for NSF (this appropriations request is colloquially known as the ``Ask''). During the authorizations and appropriations processes in Congress, the budget narratives provided by the HEP community justify the funding levels in these appropriations requests.

\subsection{P5 and HEP Community-Driven Government Engagement}
\label{sec:advocacy}

The High Energy Physics Advisory Panel (HEPAP) is a federal advisory committee that provides advice and guidance to OS HEP and NSF regarding experimental and theoretical HEP activities. Its charge \cite{HEPAP_charge} is to facilitate program reviews and long-range planning and to provide advice on funding levels. The Particle Physics Project Prioritization Panel (P5) is a sub-panel of HEPAP formed as needed to address questions about HEP projects and their planning. Recently P5 has been convened following iterations of the HEP community planning process (Snowmass) and in each case has produced a report of its findings. Of particular note are the 2013 \cite{P5-report-2013} and 2008 \cite{P5-report2008} P5 reports.

In 2013 P5 was charged with producing a strategic plan for U.S. HEP. This plan assumed a 10-year timescale in the context of a 20-year global vision for the field. The charge asked for an assessment of the (then) current and future scientific opportunities over the following 20 years, considering the field's (then) current state. The charge included three budget scenarios as reference points when forming recommendations. The full 2013 charge to P5 is available in Ref. \cite{P5-charge-2013}, and the report, which was approved in May 2014 by HEPAP, is available in Ref. \cite{P5-report-2013}. Since then, the 2014 P5 report has provided a focal point for all HEP community-driven advocacy and has factored explicitly into initial budget proposals sent to Congress and Congressional deliberations. We note that the activities of the current Snowmass process will culminate in a report expected to be utilized by HEPAP in constructing a new P5 report in 2023. This report will become the new focal point for HEP advocacy. 

\subsection{Recent HEP Funding Levels}
\label{sec:recent_funding}

\begin{figure}[htb!]
\centering
\includegraphics[width=1\linewidth]{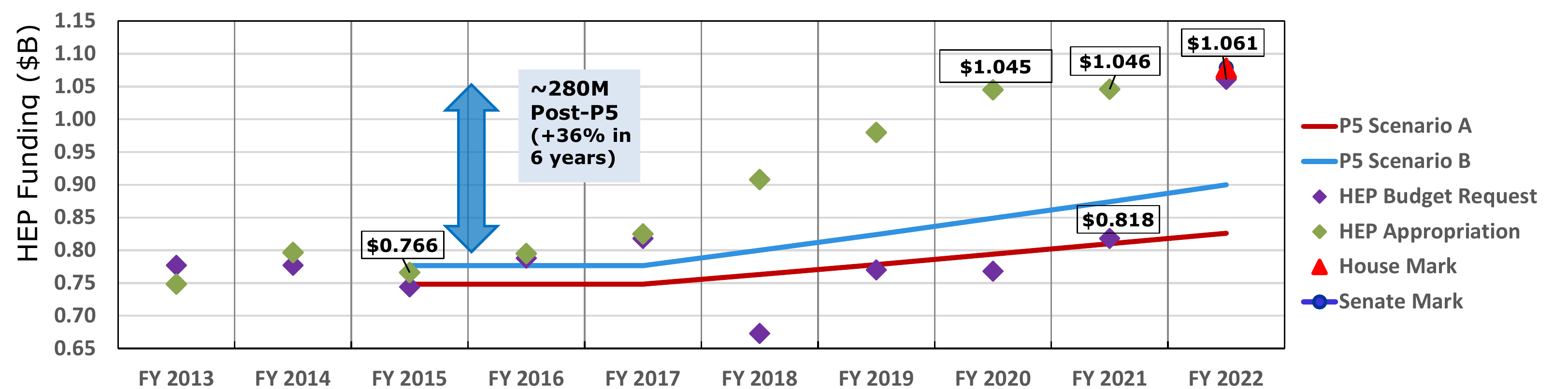}

\caption{DOE OS HEP funding since the last P5~\cite{march_hepap}. For a given year, purple denotes the PBR and green denotes the Congressional  appropriated budget.}
\label{fig:OHEP_fun}
\end{figure}

\begin{figure}[htb!]
\centering
\includegraphics[width=0.9\linewidth]{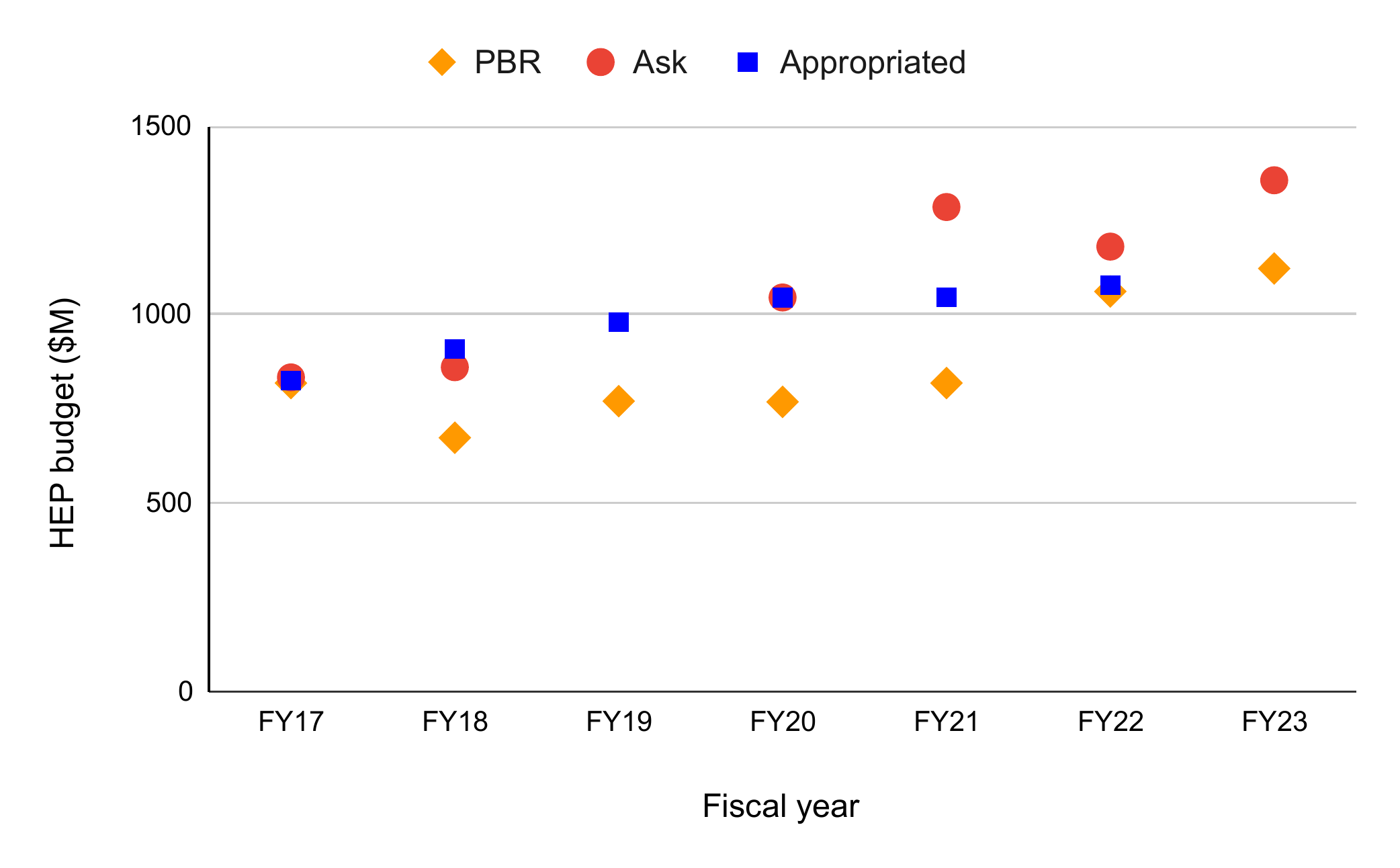}
\caption{PBR, HEP community request, and Congressional appropriated budgets in millions of dollars.}
\label{fig:budget_trends}
\end{figure}

The community's budget request and President's Budget Request (PBR) often differ, as does the amount ultimately appropriated by Congress. Figure~\ref{fig:OHEP_fun} shows the PBR and Congressional appropriated budgets, compared to the scenarios laid out in the 2014 P5 report, in recent years. Figure~\ref{fig:budget_trends} and Table~\ref{tab:request} compare the PBR, the HEP community ``Ask'', and the Congressional appropriated budgets over a similar time frame. Numbers are only shown for DOE OS HEP, as HEP is not the dominant fraction of NSF funding. Note that the impact of the 2014 P5 report on the HEP budget was not seen until 2015 due to the multi-year time frame of the budget process.

\begin{table}
    \centering
    \begin{tabular}{c|c|c|c}
    Year & President's Request (\$M) & Our Ask (\$M) & Appropriated (\$M)\\
    \hline
    FY23 & TBD   & 1356     & TBD   \\
    FY22 & 1061 & 1180     & 1078  \\
    FY21 & 818   & 1285     & 1046  \\
    FY20 & 768   & 1045     & 1045  \\
    FY19 & 770   & N/A$^{*}$& 980   \\
    FY18 & 673   & 860      & 908   \\
    FY17 & 818   & 833      & 825   \\
    \end{tabular}
    \caption{PBR, HEP community request, and Congressional appropriated budgets in millions of dollars. This is also shown in Figure 2.  {\footnotesize *Due to delays in the FY19 budget process, no advocacy was aligned with the construction of this budget.}}
    \label{tab:request}
\end{table}

\section{Accountability for Acting on CEF06 Recommendations}
\label{sec:future_group}

Existing HEP community-driven advocacy activities are broadly considered, within the community,  to be highly successful at engaging the U.S. federal legislature. These activities are jointly coordinated by the Fermilab Users Executive Committee (UEC), SLAC Users Organization (SLUO), and US-LHC Users Association (USLUA), with input from the American Physical Society Division of Particles and Fields (APS DPF) Executive Committee. Despite the longevity and success of the collaboration between these groups, this existing organizational structure does not provide sufficient resources to act on all the recommendations outlined in this report. 

The infrastructure that supports HEP community advocacy, which has grown organically over the past several decades into its current form, is structured entirely around the execution of annual advocacy activities (\textit{i.e.} the ``DC Trip''). This organization is appropriate and effective but doesn't provide a mechanism for resources to be efficiently invested into broadening the scope of community activities or achieving other long-term goals. As outlined in this report, there are several areas in which the community would benefit from expanding the scope of government engagement activities. There presently exists no group or collection of groups that could assume responsibility for directing such an effort. The authors of this report have been careful to provide actionable recommendations. Still, in several cases, progress will be constrained by the lack of a group responsible for taking action. 

The potential composition of such a group to oversee the development of long-term HEP community government engagement initiatives was a recurring theme throughout CEF06 proceedings. We concluded that it would be premature at this stage to recommend a specific solution. Instead, the groups that have historically directed HEP community government engagement activities should be charged with advancing this discussion.

\vspace{1ex}
\begin{recs}
    \rec{1}{Representatives of APS DPF, HEPAP, and the user groups, as appropriate, should have dedicated discussions to determine what actions can be taken to advance the recommendations outlined in this report.}{There is limited bandwidth in the current infrastructure for investing resources into additional government engagement activities, and the recommendations here cover a broader scope than existing community-driven advocacy. These discussions should specifically include considering the option of forming an HEP community government engagement group, composed of elected community representatives and policy experts, that is explicitly responsible for developing expanded government engagement capabilities.}
%
\end{recs}
\vspace{0.5ex}

We note that of the presently available options for an existing group to assume this role, each is constrained in key ways that would impact their efficiency. The leadership of UEC, SLUO, and USLUA are necessarily preoccupied with the execution of annual community advocacy activities. APS DPF exists within the larger structure of APS and is therefore limited in its ability to elevate, through advocacy, the interests of HEP above any other area of physics. In addition,  the community must respect the limitations placed on federal funding concerning advocacy for policy; individuals affiliated with the national laboratories or benefiting from federal grant money are subject to these limitations. 

We additionally note that the above issue exists across the whole of the Snowmass Community Engagement Frontier. In this frontier, we lack a clear analog to P5, which is charged with digesting the physics recommendations produced by the Snowmass physics Frontiers. As such, there is a greater danger that our recommendations will stagnate without clearly identified groups to act on them.
\section{Garnering Support for P5 2023}
\label{sec:community_unity}

The stakes are high for the HEP community, which currently has an annual budget in excess of USD 1B and is designing future projects that will require similar or even greater levels of funding. These budgets, ultimately derived through the U.S. federal budget process as described briefly in Section \ref{sec:funding_summary} (and in more detail in Ref.~\cite{cef06paper1}), are reaffirmed and renewed annually. Translating excitement and interest in HEP into public and federal support for funding is critical to the long-term health of the field and the larger scientific enterprise in which HEP plays an essential role. 

\begin{recs}
    \rec{2}{HEPAP must build community unity around the 2023 P5 plan and develop a clear messaging strategy spanning the next 10 years. 
}{Impactful community-driven advocacy and strong support for HEP within Congress are dependent on a unified community message that reflects the priorities outlined in the P5 report. Dedicated efforts to engage the HEP community and all relevant stakeholders are required to build and maintain this unity as the 2023 P5 report is developed, rolled-out, and implemented.}

\end{recs}

One of the fundamental aspects of the P5 plan and report is that they outline a community-wide strategy for \textit{all} HEP that can garner broad support and community buy-in, thereby resulting in a singular community-wide message about the status and future of our field. This community unity around a singular plan has been seen as necessary by policymakers, as has the successful implementation of the plan. The success since P5 2014 of this unified messaging is in notable contrast to the state of community messaging before 2014, which was significantly more fragmented. Of particular relevance for maintaining the unity of our messaging is the science advocacy undertaken by individuals outside of community-organized advocacy. It is critical that the messaging from these individuals be consistent with HEP community messaging. The HEP community collectively has a responsibility to communicate the community message broadly and to provide the knowledge and supporting material to convince everyone inside and outside the community that the P5 plan is worthy of support.

Actions that should be taken to ensure the maximal impact of the P5 plan can be divided into two time frames. First are short-term actions pertaining to the drafting of the P5 report and its initial roll-out. Second are long-term actions that should be taken over the next decade to ensure that the plan remains relevant as budgetary, scientific, and other circumstances may evolve. In addition to these, it is critical that action be taken to begin implementing the recommendations of the Snowmass Community Engagement Frontier, which in many areas highlight systemic barriers to building community unity.

\subsection{Short-Term Actions to Build Unity}
\label{sec:unity-short-term}

HEPAP and the 2023 P5 panel must lead the effort to build community support for the new P5 plan. Failure to sufficiently educate and convince the community, Congress, and the executive branch about the merits of the 2023 P5 plan jeopardizes the successful implementation and execution of projects outlined in the plan. The first aspect of this approach should be a P5 ``feedback campaign'', including opportunities for feedback, detailed question and answer sessions, colloquia, seminars, and conference talks. In order for the community to rally behind the P5 report, sufficient opportunity must be presented for all voices to be heard directly as the document is drafted. The Snowmass and HEP communities should have explicit opportunities to provide feedback outside of HEPAP meetings before the P5 report is finalized.

\begin{recs}
    \subrec{2.1}{HEPAP and P5 should solicit feedback on the 2023 P5 report from HEP community members and all other relevant stakeholders before it is finalized.}
\end{recs}

Once the P5 report is complete, HEPAP and P5 should undertake an ``education campaign'' to broadly communicate the details of the finalized report. Meetings should be arranged with all relevant stakeholders, including national laboratories, domestic and international funding agencies, the Executive Office of the President (OMB and OSTP), and experimental collaborations 

\begin{recs}
    \subrec{2.2}{HEPAP should implement a wide-reaching outreach program to communicate the 2023 P5 plan to the HEP community and all other relevant stakeholders. This program should include education on the details of the plan and emphasize the importance of clear messaging on the priorities.}

\end{recs}

\subsection{Long-Term Actions to Maintain Unity}
\label{sec:unity-long-term}

In the years since the publication of the 2014 P5 plan, the P5 chair has led an annual effort to produce status reports that summarize the community's progress with respect to its execution~\cite{uspp_pp}. This effort has been a highly successful element of centering the P5 plan in community-driven outreach activities targeted at Congress. It is vital that this effort be continued, ideally by the 2023 P5 chair, and supported until the next P5.

\begin{recs}
    \subrec{2.3}{The P5 chair or equivalent should continue to produce the one-pager (or equivalent) status reports that summarize the community's progress in executing the 2023 P5 plan annually.}
\end{recs}

Though status updates on P5 progress and priorities are presented at HEPAP and similar venues, feedback provided by community members in the course of CEF06 proceedings indicates that these communications are not generally disseminated to the broader HEP community. Prioritizing regular talks at APS, DPF, and other HEP-wide conferences would help to keep the community engaged and informed in the coming years as the 2023 P5 plan is executed. 

\begin{recs}
    \subrec{2.4}{HEPAP should provide regular status updates on P5 progress and priorities broadly within the HEP community through mechanisms such as talks at APS, DPF, and other HEP-wide conferences to keep the community engaged and informed as the 2023 P5 plan is executed.}
\end{recs}

Since the last P5, we have seen natural deviations from the plan laid out in the 2014 P5 report due to budgetary changes, physics discoveries, the development of new technologies, and shifts in Administration priorities. Similar deviations should be expected throughout the execution of the 2023 P5 plan, forcing the funding agencies to choose how best to accommodate them. A mechanism must be created to communicate such changes and deviations from the P5 plan to the community as they arise. Specifically, these should be communicated and explained to the community through written reports and conference presentations. Additionally, mechanisms should be created to facilitate community input into such decisions.

\begin{recs}
    \subrec{2.5}{HEPAP must communicate to the community and provide avenues for feedback when there are changes to the implementation of the 2023 P5 plan that arise due to evolving funding profiles, changes in Administration or HEP priorities, or other external circumstances.}
\end{recs}

In addition to maintaining the infrastructure to respond to external circumstances over the next decade, the community would be well-served by developing the infrastructure to monitor internal circumstances, particularly in areas where there is overlap with the activities outlined by P5. This should include the systematic collection of data on an annual basis such as the fraction of the HEP portfolio designated for research, and the size, career breakdown, and diversity of community membership. Having these data available will help to understand the full impact of the 2023 P5 plan over time as it is implemented.

\begin{recs}
    \subrec{2.6}{HEPAP should continuously assess the impact of the 2023 P5 plan on the community over the course of its implementation. This should include providing mechanisms for community feedback on the impact of the plan and monitoring relevant community metrics. Suggestions for metrics include the fraction of the HEP funding portfolio allocated to research and the size, growth, and diversity of the field.}
\end{recs}
\vspace{0.5ex}

\subsection{Progress is Required on Community Engagement Topics}
\label{sec:cef-topics}

Tremendous progress has been made during this Snowmass cycle in the area of ``community engagement'' topics. The activities of the Community Engagement Frontier (CEF) were not focused on questions about \textit{the science} that our community engages in, but rather were focused on questions about \textit{how we do science}. Conversely, the P5 process is designed to produce a roadmap for the particular science (organized into projects) that our community will undertake in the coming years, but not necessarily to weigh in on the many factors that will impact the ability of community members to engage in that work. 

There is no precedent for the P5 report to include any ``community engagement'' guidance or recommendations. And it is not clear that the P5 scope should be expanded to include consideration of such topics. However, it is clear that many of these topics are critical to the foundation of the HEP community and there are many areas in which action must be taken to ensure the future success of the field. Of particular concern is the talent that the community stands to lose if progress is not made in areas that the early-career generation of community members view to be highly important. At their core the issues identified within CEF collectively represent a barrier to community unity, and inaction in this area poses a severe threat the to the successful implementation of the 2023 P5 plan.

\begin{recs}
    \subrec{2.7}{DOE and NSF should include CEF topics in the P5 charge or create a separate HEPAP subpanel focused on CEF topics. Not all topics within the purview of Snowmass CEF may be appropriate for such a HEPAP subpanel, but this would provide an important mechanism for specific community goals to be defined and our progress towards achieving them to be tracked.}
\end{recs}
\section{Congressional Advocacy for HEP Funding}   
\label{sec:current_advocacy}

Community-driven advocacy is critical to sustaining strong support for HEP funding and priorities. Historically (for the past 30+ years), this advocacy has been centered on an annual trip to Washington, D.C. by a delegation of HEP community members. The principal goal of this effort is for the delegation to visit as many Congressional offices as possible. In meetings with Members of Congress and their staff,  community members share their excitement about HEP research and solicit support for HEP from policymakers. This effort also includes meetings with executive branch offices (OMB and OSTP) and representatives of the funding agencies (DOE OS, DOE OS HEP, and NSF).

The goal of this effort is to cultivate support for the funding of physical sciences research in general and HEP in particular. Community members discuss the overall benefits of HEP and basic research with policymakers and deliver an appropriations ``Ask'' for HEP funding. The ``Ask'' includes a request for specific funding levels for each of DOE OS HEP and NSF. The timing of the trip is chosen to align with the annual budgetary cycle when the budget is being discussed by Congress (typically this is in March).

The percentage of offices visited has grown over the years as support for the trip has grown. In 2010, 34 community members participated; by 2017 this number rose to 50. In 2017 about 70\% of House and Senate offices were visited. Following an increase in funding that enabled greater participation, in 2019 100\% of House and Senate offices were visited, and the number of trip attendees approached 70. Coordinating the meetings of 70 community members with up to 541\footnote{Including six non-voting members of the House of Representatives.} Congressional offices requires considerable organization. Over the years, several sophisticated tools have been developed to support these efforts. 

The annual effort is jointly coordinated by the Fermilab Users Executive Committee (UEC), SLAC Users Organization (SLUO), and U.S.-LHC Users Association (USLUA), with input from the American Physical Society Division of Particles and Fields (APS DPF) Executive Committee. The primary affiliations of trip attendees in recent years have been 50\%, 35\%, and 15\% with UEC, USLUA, and SLUO, respectively. Attempts are made to recruit participants from different career stages, demographics, and research areas, though the organization of the trip is limited to the user groups. Historically, the chair of the UEC Government Relations subcommittee has assumed the responsibility of leading this organization. 

Through election, the groups responsible for the organization of this effort represent a large portion of the U.S. HEP community, but not the community as a whole. There are constituencies within the HEP community that are systematically excluded by focusing the organization within national laboratory user groups. This was a topic of discussion during the proceedings of CEF06, and though we did not arrive at a specific conclusion, we recommend that this conversation be continued. In particular, the community should consider the creation of a new group with elected membership to assume the responsibility for organizing this effort that can be explicitly representative of the entire HEP community.

\begin{recs}
    \rec{3}{Representatives of UEC, SLUO, USLUA, and APS DPF should facilitate discussions to consider the formation of a more formal ``HEP Congressional advocacy'' group to assume responsibility for organizing the annual advocacy trip to Washington, D.C.}{The formalization of such a group has the potential to streamline the adoption of other recommendations in this report and to elevate the profile of this important effort within the HEP community. Any discussions on this topic should include other leaders of the HEP community and provide mechanisms for feedback from community members.}
\end{recs}

Throughout this document we will use the term ``HEP Congressional advocacy group'' to refer to the group responsible for organizing annual HEP community-driven advocacy activities targeted at Congress, in its present form or any hypothetical future form. The remainder of this section describes specific aspects of the present organization of annual community-driven advocacy and provides specific recommendations for how these activities can be improved. The collection of these specific recommendations can be summarized by the more general recommendation below that this important work needs to be sustained.

\begin{recs}
    \rec{4}{The HEP Congressional advocacy group must continue to support, and should aim to grow, the annual HEP community-driven advocacy activities.}{The annual community-driven advocacy effort is essential to increase knowledge and interest of HEP in Congress. Participation in these activities should be encouraged within the community, and community leaders should support efforts for continued development and growth of this effort.}
\end{recs}

\subsection{HEP Advocacy Tools and Logistics}
\label{sec:tools}

\subsubsection{The Washington-HEP Integrated Planning System (WHIPS)}
\label{sec:whips}

The Washington-HEP Integrated Planning System (WHIPS) is a framework developed to handle most of the logistics of planning, executing, and documenting HEP community advocacy activities. Aspects of the trip planning logistics were developed by multiple individuals over many years, and a considerable effort was undertaken in 2018 to consolidate these aspects into a unified framework. WHIPS has significantly improved the efficiency of organizing, and thereby the success of, the annual HEP advocacy effort. The functions that WHIPS efficiently serve were previously handled manually, involving significant last-minute work by the trip organizers. The development and maintenance of WHIPS has been a voluntary activity by a small group of early-career community members. Single points of failure and the lack of a clear plan to facilitate knowledge transfer risk its long-term utility.

WHIPS stores information about trip attendees and their connections to states and Congressional districts, details of Congressional offices and (sub)committee assignments, and all future and past meetings between trip attendees and Congressional staff. An algorithm has been developed that assigns trip attendees to Congressional districts based on where the attendees have lived, worked, voted, or have family. These connections between a community member and a Member of Congress enable more focused meetings that cater to that district's or state's priorities and in many cases provide a foot in the door for setting up meetings. WHIPS has significantly contributed to the recent success of the advocacy effort, in particular providing technical capability to support the current number of trip participants and to hit the target of visiting 100\% of offices in 2019~\cite{whips-talk}.

We strongly endorse the continued utilization of this valuable resource, and recommend that steps be taken to ensure that its maintenance and administration be supported in future years. Specifically, we note that the past and present developers of WHIPS are all early-career members of the community without permanent (long-term) positions -- this presents the potential for expertise to become unavailable if developers move out of the field as their careers progress. We note that during the proceedings of CEF06 it was suggested that community members with permanent positions could be intentionally brought into the development team to get involved in the maintenance and administration. Along the same line of reasoning, it has been proposed that the administration and development of WHIPS be formally brought under the ownership of one of the organizations that has historically led community advocacy activities (or a future organization created more specifically for the purpose of leading advocacy efforts). 

\begin{recs}
    \subrec{4.1}{The HEP Congressional advocacy group must ensure continued access to and sustained maintenance of WHIPS, which has been enormously additive to the annual advocacy efforts undertaken by the HEP community.} 
\end{recs}

We additionally note that the developers of WHIPS maintain a list of technical features and functionalities that have been proposed by users and we observe that action on these has historically been focused on the imminent needs of a particular year's advocacy efforts. The HEP Congressional advocacy group should allocate labor and resources to continue the development of WHIPS between advocacy cycles to better accommodate less-urgent but still worthwhile development tasks.

\subsubsection{Other Advocacy Tools}
\label{sec:other_tools}

In addition to WHIPS, the annual advocacy effort has historically utilized several other technologies to organize and execute activities each year. Two of particular note are the ``Twiki'' page\footnote{We note that this is not a publicly accessible resource, and therefore do not provide a reference to it.}, maintained by the Fermilab UEC, and the HEP funding website and grants database \cite{baumer}. Specific ideas which have been raised for how to expand the utility of the grants database include tracking HEP grants by Congressional district and developing improved metrics to track the fraction of NSF funding that is used for HEP-specific projects and grants. The HEP Congressional advocacy group should establish a long-term plan for the continued utilization of these tools. This plan should include contingencies if underlying technologies supporting these resources become unavailable. We also note that progress has been made recently in importing aspects of the grants database into WHIPS.

\begin{recs}
    \subrec{4.2}{The HEP Congressional advocacy group should establish a long-term plan for the continued utilization of all technical tools that support HEP community advocacy activities, including the ``Twiki'' and HEP funding website and grants database.}
\end{recs}

\subsubsection{Internal Advocacy Diagnostics}
\label{sec:diagnostics}

We strongly endorse the development of improved tracking diagnostics to understand the efficacy of HEP community advocacy activities. The advent of WHIPS has made this much more technically feasible, and several specific diagnostics are presently built into the WHIPS database system (\textit{e.g.} tracking the advocacy activities of individual community members and for individual Congressional offices). However, we note that studies of such metrics are, by definition, outside the scope of the planning for advocacy activities in any individual year due to the long time scale on which their analysis would yield valuable data. Therefore, a dedicated conversation should be undertaken by the HEP Congressional advocacy group on this topic, and an intentional effort should be made to expand the tracking of diagnostic information within WHIPS or elsewhere. Examples of diagnostic information pertaining to the offices visited that could be tracked in WHIPS include the voting records of individual offices or Members of Congress, appropriations requests delivered to offices, and signatories of the ``Dear Colleague'' letters. Also, diagnostic information pertaining to the subset of the HEP community that engages in advocacy should be developed -- this suggestion is expanded on in Section~\ref{sec:representation}.

\begin{recs}
    \subrec{4.3}{The HEP Congressional advocacy group should facilitate a conversation about advocacy diagnostics that would enable an analysis of the long-term impact of HEP community advocacy and should identify a plan for collecting relevant data.}
\end{recs}

\subsection{Advocacy Training}
\label{sec:advocacy_training}

Community members participating in HEP advocacy activities have varying \textit{a priori} levels of understanding of the U.S. budget and appropriations process, how to communicate science, and how to interact with policymakers. They need to be trained in these areas and in the day-to-day and meeting-specific logistics of the advocacy trip. Over many years, training material has been developed to educate participants on each topic quickly. In advance of the trip each year several dedicated sessions are held to train trip attendees on topics including: an overview of the U.S. budget process; how to efficiently communicate about science; the specifics of that year's ``Ask''; WHIPS; how to schedule meetings; and how to effectively participate in meetings. In recent years veteran trip attendees have produced example videos showing how to conduct meetings. In aggregate, these recordings cover a variety of styles and scenarios reflective of the spread in how these meetings can take place. 

These advocacy training resources have been critical to the recent success of community-driven HEP advocacy activities and resources should continue to be invested to support them. Additionally, a subset of these resources would be beneficial to the broader HEP community for facilitating discussions between community members and the general public about the societal impacts of our field. Making these materials more broadly available within the community would additionally have the effect of raising awareness within the community of HEP advocacy activities.

\begin{recs}
    \subrec{4.4}{The HEP Congressional advocacy group must ensure that resources continue to be allocated to provide advocacy training to ``DC Trip'' attendees. In addition, relevant parts of the training currently used for this effort should be made more widely available to the community to increase the community's policy literacy overall, independently of the goals of any year's particular advocacy efforts.}
\end{recs}

\subsection{Community Advocacy Inreach}
\label{sec:community_inreach}

The planning and execution of the annual community advocacy activities are generally contained within a small subset ($\sim$100) of the community composed mainly of members of the executive leadership of the user organizations (Fermilab UEC, SLUO, and USLUA) and DPF, and some past trip attendees. We strongly suggest that having more knowledge and discussion of the activities in various formats would benefit the field and positively impact our advocacy goals. For example, broader inreach into the community would help us to achieve our goals to have a more representative group participate in advocacy and to build stronger connections to every district. The current state of community inreach about community-driven advocacy activities is minimal. There are a limited number of updates provided to the community each year at APS DPF and user group annual meetings, as well as yearly updates provided to HEPAP at public meetings (beginning in 2017, see Refs \cite{hepap_2017, hepap_2018, hepap_2019, hepap_2020, hepap_2021}). Additionally, some details are available in user group meeting minutes, reports, and web pages. However, the overall reach of these means of communication into the community is limited, and awareness of advocacy efforts is generally deficient.

We recommend that a summary of annual advocacy activities regularly be assembled in a document suitable for distribution within the community. We note that a high-level executive summary \textit{is} presented each year to HEPAP, though its existence is not effectively broadcast within the community and it is likely inaccessible to members of the community that lack the public policy expertise of HEPAP. We additionally note that specific details of the trip may not be suitable for consumption outside of the community (\textit{e.g.} themes of interactions with the offices of current Members of Congress), but believe that these details can be omitted with due care taken in the preparation of a summary. The goal of such a summary should be to increase awareness of advocacy efforts in the community and to cultivate interest in participation in future advocacy activities.

\begin{recs}
    \subrec{4.5}{The HEP Congressional advocacy group should enhance efforts to educate the HEP community about advocacy activities. This effort should include invited and submitted talks at conferences, collaboration meetings, and other community gatherings. The goal should be increasing awareness of the trip and providing expanded opportunities for community members to participate in organizing and engaging in advocacy activities. An annual report summarizing community advocacy activities should be produced and distributed widely within the community.}
\end{recs}

\subsection{Continuity and Succession of Trip Leadership}

The leadership of the annual advocacy trip has historically been provided by the Fermilab UEC, SLUO, and USLUA, and has been directed by the chair of the UEC Government Relations subcommittee. This model has its strengths, namely in enabling the rotation of the time-intensive coordination of logistical aspects of the effort among community members annually while simultaneously providing the basis for institutional knowledge to be preserved within the executive leadership of the three user groups. There are many nuances to planning the annual advocacy effort which require the expertise of community members who have repeatedly participated in this effort over many years and have developed considerable experience. Examples of this nuance include sustained long-term relationships with Congressional subcommittee staff and other elected officials, logistical expertise, and experience communicating legislative details with career staff members at OMB and OSTP.

It has been noted repeatedly by community members that the model in use is susceptible to single points of failure in the form of individuals relied upon for a disproportionate amount of subject matter expertise. Specifically, it has been suggested that a plan be developed to intentionally expand the pool of community members that have specialized knowledge in the following key advocacy areas:

\begin{itemize}
    \item Organizing meetings with funding agencies and executive branch agencies during the annual trip to Washington, D.C.;
    \item Organizing and conducting meetings with (sub)committee staff during the annual trip to Washington, D.C.; and
    \item Development and administration of WHIPS and other technical elements of the infrastructure that enables HEP community advocacy.
\end{itemize}

We recommend that the executive leadership of the user groups and DPF have dedicated discussions to assess the robustness of advocacy leadership continuity and identify steps that can be taken to mitigate points of failure. 

\begin{recs}
    \subrec{4.6}{The HEP Congressional advocacy group should prioritize the creation of documentation to guide future advocacy effort leaders in the planning of annual activities and this documentation should be made broadly accessible within the community.}
\end{recs}

\subsection{Improving Representation in Advocacy Activities}
\label{sec:representation}

The groups historically represented by the organizers of annual advocacy activities provide broad, but not complete, academic representation for the U.S. HEP community. For example, theorists make up a smaller fraction of the users community than the whole HEP community because the national laboratories focus primarily on experimental efforts. It is also the case that most participants in the annual advocacy efforts have historically been members of the Energy and Intensity\footnote{We note that the Intensity Frontier, as defined in the proceedings of Snowmass 2013, maps nearly onto the Neutrino and Rare Process Frontiers, as defined in the proceedings of Snowmass 2021.} Frontiers, with smaller representation from the Theory, Cosmic, Computational, and Instrumentation Frontiers. There is no comprehensive demographic information available for participation in annual advocacy activities, especially farther back than recent years. Still, there is \textit{some} information available from post-trip surveys in recent years which supports these assertions.

According to the last few years of data, approximately one-third of participants have been postdocs, one-third professors or national laboratory scientists, and the final third has been distributed between Ph.D. students and other university or laboratory staff. It is expected (and indeed intended) that the demographics of the group will skew towards early career members of the community, but more could be done to increase the representation of graduate students in particular.

Historically, participation by underrepresented and marginalized groups in the advocacy efforts has not been explicitly tracked. The users groups must take actions to quantify and improve representation through the invitation of individuals or by working with groups representing underrepresented and marginalized demographics in HEP.

\begin{recs}
    \subrec{4.7}{The HEP Congressional advocacy group should track the demographics of individuals engaging in community-driven advocacy activities, and efforts must be made for the demographics of participants in these activities to be representative of the field.}
\end{recs}

We note that the creation of a dedicated HEP Congressional advocacy group with directly elected membership, as suggested in Recommendation 3, could provide an avenue to more completely reflect the demographics of the field. This would be particularly relevant to broadening the participation of community members based at smaller institutions, or who are otherwise unaffiliated with the national laboratories.

\subsection{Expanded Time Frame for Community Advocacy} 

The majority of existing HEP community advocacy activities center on an annual trip to Washington, D.C. in the Spring, generally timed to coincide with the drafting of appropriations and authorizations bills at the subcommittee level in Congress. There is some precedent for additional, targeted, in-person advocacy in Washington, D.C. by a smaller delegation. In these cases, the trip is timed to coincide with debate surrounding the final details of the budget, generally in the Fall. During the proceedings of CEF06 community members repeatedly observed that there is the potential to expand the scope of advocacy efforts to regularly include such ``off-cycle'' activities. These off-cycle activities could also include, for example, trips to district offices (rather than to Washington, D.C. offices) in the Fall to reinforce the messaging delivered each Spring.

Expanding the time frame in which the community engages in advocacy would provide the opportunity to boost participation (\textit{i.e.} increased participation by a broader subset of the community), albeit at the expense of additional resources required on the organizational/management side. We note that such an effort would have been prohibitively resource-intensive before the advent of WHIPS and that WHIPS potentially lowers the threshold for introducing additional advocacy activities to an acceptable level. One idea which has been suggested is to create a new profile tier within WHIPS to provide community members with the resources (\textit{e.g.} automatically generated letters) to engage in targeted advocacy of their local (Congressional) offices.

It is worth carefully considering the marginal increase in resources required to sustain additional advocacy efforts throughout the year beyond what is already required to support the current advocacy model. We do not offer a recommendation as to whether such a model of more continuous advocacy is appropriate. However, we recommend that a conversation around this topic be facilitated with more careful consideration of the potential costs and benefits.

\begin{recs}
    \subrec{4.8}{The HEP Congressional advocacy group should investigate the potential for year-round advocacy as an extension of annual HEP Congressional advocacy activities.}

\end{recs}

\section{HEP Communication and Outreach Materials} 
\label{sec:outreach_materials}

A vital resource for HEP community-driven advocacy activities is a series of informational pamphlets designed to concisely relay the community's most important messages in a visually appealing and accessible manner. These materials, which are available on the U.S. particle physics website \cite{uspp2022mar} are jointly maintained by APS DPF, UEC, SLUO, and USLUA. In addition to serving as an important resource for advocacy activities, they serve the larger purpose of communicating about HEP and its benefits to a general audience, and they have multiple applications of which advocacy activities are just one.

These community communication and outreach materials are essential to the Congressional HEP community-driven advocacy strategy but are separate from it in key respects. During the HEP Congressional visits, these materials serve a dual purpose: to provide a conversation piece during meetings and to leave a reference guide for Congressional staff as they make their recommendations following our meetings with them. We note that due to the Hatch Act of 1939, government employees are expressly barred from engaging in political activities, so these materials have always been \textit{communication} material and explicitly do not advocate for any area of government support. Their purpose is to inform non-community-members in general about HEP and its benefits.

A committee of volunteers is formed to produce this material and update it on an annual basis. This committee is composed of representatives from the UEC, USLUA, and SLUO, with support from the Fermilab Office of Communication, DOE, and the former P5 chair. Recently, this effort has been supported by a member of DOE who has helped the community coordinate the drafting of these materials and has served as a repository of knowledge and as the group leader, ensuring that steady progress is made on schedule. We note that the contributions of various individuals have been extremely helpful in crafting the trip materials but that there is no formal requirement that this support is provided by the Fermilab Office of Communication, DOE, or by the former P5 chair. There is a risk that this support may someday end, and it would behoove the community to identify a way of maintaining records and support from these organizations and individuals.

Below is a list of the documents produced through this effort and the aspect of HEP that each seeks to explain.

\begin{itemize}
    \item \textbf{Particle Physics Progress and Priorities.} Annually updated, this document examines the HEP community's recent accomplishments and the priorities for the upcoming year within the focus of the P5 plan.
    \item \textbf{Particle Physics is Discovery Science.} This document discusses the broad questions the HEP community is trying to answer and how they tie back to big questions called out in the 2014 P5 report.
    \item \textbf{Particle Physics Makes a Difference in Your Life.} This document discusses some ways in which HEP has impacted other fields and industries.
    \item \textbf{Particle Physics Builds STEM Leaders.} This document discusses the outreach and public engagement activities of the HEP community.
    \item \textbf{Particle Physicists Value Diversity and Strive Toward Equity.} This document contains a statement of HEP community values, describes current diversity, equity, and inclusion goals, and highlights ongoing programs designed to empower and provide opportunities for historically underrepresented and marginalized groups.
    \item \textbf{Particle Physicists Deliver Discovery Science Through Collaboration.} This document shows that HEP is an international effort involving some of the best minds worldwide. It also summarizes the plans, timelines, and present status of projects identified in the 2014 P5 report.
    \item \textbf{Particle Physicists Advance Artificial Intelligence.} This document explains how the HEP community uses machine learning (ML) and successfully interfaces with industrial partners to push the boundaries of ML research and development.
    \item \textbf{Particle Physics and Quantum Information Science.} This document discusses the benefits of QIS, the skill sets that make particle physicists valuable in these endeavors, and how QIS developments can, in turn, help solve fundamental problems in our science.
    \item \textbf{Particle Physics in the U.S. Map.} This map shows the distribution of institutions involved in HEP that receive funding from either DOE or NSF.
\end{itemize}

\begin{recs}
    \rec{5}{The HEP Congressional advocacy group must continue support for the development and maintenance of HEP Communication materials, working with the 2023 P5 chair and other experts as needed.}{High quality and well-developed communication and outreach materials are essential for effective government outreach, and their quality reflects directly on how our field is perceived. Investment in these important resources must be continued, and they should be made more broadly available within the community.}
\end{recs}
\section{Community Advocacy on Topics Beyond HEP Funding} 
\label{sec:non_funding_topics}

The existing framework for HEP community-driven advocacy is focused on cultivating general support for and awareness of basic research and HEP activities within Congress, with the explicit purpose of maintaining and growing the funding of the field. As discussed in Section \ref{sec:outreach_materials}, a vital element of this effort is fostering discussions about non-funding topics (\textit{e.g.} the impact that HEP developments have on society). Still, these are discussed only to the extent that they directly support the budget-oriented priorities of targeted advocacy activities.

Many topics beyond federal funding significantly impact HEP research and researchers. Throughout the proceedings of CEF06, extensive discussions took place about how the community might organize or participate in other government engagement for such topics which are not directly tied to the funding of the field. In this section, we briefly introduce a small number of these topics, discuss how these topics sometimes arise in conversations with Congressional offices in the context of existing (funding) advocacy and highlight external resources available to community members to engage in meaningful advocacy in this area. Additional details on many of these topics are available in the CEF03 (Diversity \& Inclusion) topical group report \cite{cef03}.

\subsection{Legislative Topics Impacting HEP}
\label{sec:non_funding_issues}

Below are examples of topics that directly impact the HEP community and individuals within the community. These examples all share the potential to be significantly affected by policy actions of the U.S. federal legislature. We note that this list is not exhaustive and that more details about these topics are available in Refs.~\cite{cef06paper2} and~\cite{cef03}.

\textbf{``Diversity, equity, and inclusion'' (DEI)} has become a catch-all term for a broad array of issues that marginalized communities experience which impact every aspect of HEP research and all HEP researchers. One example of a policy change within this area is expanded access to resources to perform research provided by the America Creating Opportunities for Manufacturing, Pre-Eminence in Technology, and Economic Strength (COMPETES) Act. Many small colleges and universities have historically had limited opportunities to participate in research. This group includes Historically Black Colleges and Universities (HBCUs), other minority-serving institutions (MSIs), and non-R-1 and smaller institutions, such as liberal-arts colleges and others without graduate programs. The COMPETES Act, recently passed by the U.S. House of Representatives, includes a  program to increase funding to such small institutions~\cite{Ambrose2022feb}.

\textbf{U.S. immigration policy} is a vital issue because of the international nature of HEP. Many scientists come to the U.S. to pursue education or academic positions, collaborate with U.S. scientists, and use U.S. research facilities. Many U.S. scientists similarly work and perform research outside the U.S. An example of an issue in this area which can be addressed through policy is the uncertainty and opacity inherent in the current U.S. immigration system which imposes substantial physical and psychological costs on HEP researchers and reduces the field's efficiency and productivity. In addition, the widespread practice of educating Ph.D. researchers in the U.S. while providing few paths for visas or permanent residency for graduates afterward harms national competitiveness in R\&D. An example of policy targeting this issue is the Keep STEM Talent Act~\cite{Foster2019}. This policy makes individuals who earned master's or Ph.D. degrees in STEM fields in the U.S. automatically eligible to apply for permanent residency in the U.S. A recent APS report \cite{APS_intl} discusses the of international researchers on U.S. science. 

\textbf{``Research security''} is a concept that has experienced increasing emphasis recently, with largely detrimental consequences. While the need for extreme care and caution is widely acknowledged to protect intellectual propriety, basic research has always benefited from openness and collaboration, including throughout the Cold War. The recent introduction of  ``sensitive'' areas of research and corresponding concerns about ``foreign talent recruitment programs'' have been seen to have a widely-reported negative effect on international collaboration, especially with Chinese institutions and researchers~\cite{APS2021,Thomas2022jan}.

\textbf{``National laboratories access''} refers to physical access to the U.S. National laboratories as well as access to their resources, and discussions about restrictions to such access have increased in recent years. U.S. national laboratories host substantial HEP experimental resources and infrastructure that are critical for research. Even laboratories nominally open to the public often have stricter access restrictions than other research institutions, such as universities. Individuals who originate from ``countries of concern'' may be barred from accessing otherwise publicly available resources based on a background check for which there is limited transparency and no appeal. Researchers visiting for workshops or conferences may be denied visas without transparency or appeal. Denying these visas harms those researchers and U.S. HEP as a whole: critical research may be delayed or even abandoned, and the international community is increasingly viewing the U.S. as an inhospitable location for hosting important meetings. 

\textbf{``Basic science funding''} and federal spending on R\&D has been decreasing as a percentage of GDP for four decades, a trend which has not been observed globally. Therefore, although HEP has seen an increase in funding in recent years, science spending in the U.S., despite growing in absolute terms, has not kept up with economic growth or with international trends \cite{niskanen, nature}.
Numerous studies have calculated positive rates of return from basic research ranging from 20\% to nearly 100\%, depending on various economic assumptions~\cite{SALTER2001509,summers}, which indicate that the U.S. could realize substantial economic and social gains by increasing science funding well above present levels.

\subsection{Current Implicit Advocacy}
\label{sec:implicit_advocacy}

Some of the above issues, plus others, are touched on in the materials that the community uses in its current government engagement (described in Section \ref{sec:outreach_materials}). While materials that touch on these issues are part of the more extensive discussion about the benefits of supporting particle physics, existing community advocacy activities do not include advocacy for specific changes or policies in these areas. The materials and the message about topics other than HEP funding are crafted by a small group of volunteers connected to the annual Congressional advocacy effort. It has been noted (see Ref. \cite{cef06paper1}) that this group is not fully representative of our field and does not have a mechanism for broader community input to be incorporated into these messages. These topics all fall within a category of issues with general community support, but the question remains if more expert input or explicit consensus-building is needed or may be needed in the case of future issues.

The following items are topics that are implicitly advocated for as a core component of existing community advocacy activities. The specific document that discusses each topic follows the topics in parentheses. We note that this is not an exhaustive list.

\begin{itemize}
    \item STEM education (\textit{Particle Physics Builds STEM Leaders}).
    \item International collaboration; international and open science (\textit{Particle Physicists Deliver Discovery Science Through Collaboration}).
    \item DEI activities (\textit{Particle Physicists Value Diversity and Strive Toward Equity}).
    \item Emerging technologies; QIS and AI (\textit{Particle Physics and Quantum Information Science}, \textit{Particle Physicists Advance Artificial Intelligence}).
    \item Basic research support as a driver of U.S. innovation, economic development, and national security (\textit{Particle Physics Makes a Difference in Your Life}).
\end{itemize}

\subsection{External Resources for Non-budgetary Advocacy}
\label{sec:external_resources}

While the HEP community does not presently engage in advocacy on any topic apart from the HEP budget, external resources exist within the larger research community to facilitate government engagement on these topics. These groups, which represent the broader physics and science communities, have considerably greater resources than the HEP community alone. Below we summarize some physics-specific groups and point out relevant considerations for understanding how their resources could benefit the HEP community.

\textbf{The American Physical Society (APS)} is a nonprofit membership organization representing physicists within the U.S. The mission of this group is to advance and disseminate physics knowledge and advocate for the needs of physicists and scientists at large. APS advocates for many non-funding legislative issues of importance to the physics community. However, the HEP community is only a subset of the broader APS membership. APS is a trade organization available to all physicists, but the annual fee may limit participation. 
APS also offers Congressional Science Fellowships, which aid Congress by providing scientifically literate, skilled personnel. Raising awareness of this program in the HEP community could be beneficial.

\textbf{The American Institute of Physics (AIP)} is another organization that works to promote and advance the physical sciences. AIP is an umbrella organization that pools the resources of 10 member societies, including APS. As an umbrella organization, AIP can help coordinate the individual members' activities and messages. AIP also provides a Congressional staff program, very similar to the one provided by APS. AIP, too, should be promoted as a resource within the HEP community.

\textbf{The American Association for the Advancement of Science (AAAS)} is mostly known through its journal \textit{Science} and policy fellowship program. Its goal of advancing science means that it has an active government engagement group which provides training programs and has significant resources. AAAS hosts various groups to get scientists involved in policy, such as the Local Science Engagement Network and National Science Policy Network. AAAS is an underutilized resource within the HEP community, even more so than APS and AIP. Details are available on the AAAS website \cite{AAAS}.

\subsection{Taking Action on Topics beyond HEP funding}
\label{sec:non_funding_topics_action}

The examples outlined above and others are widely considered within the HEP community to be important. However, there exists no consensus on whether action should be taken by the community to influence policy in these area, let alone what specific actions would be appropriate. During the proceedings of CEF06, some individual community members advocated that existing HEP community advocacy resources be leveraged to undertake an expanded scope of advocacy activities. Other individual community members cautioned that engaging in such activities would lead to repercussions that impact the entire field. The specifics of these conversations are elaborated on in Ref.~\cite{cef06paper2}. At present, we recommend that this conversation be continued and expanded upon. Additionally, we recommend that the minimal available action be taken, which would be to direct community members to available external resources so that they are better equipped to participate in advocacy in this area if they choose.

\begin{recs}
    \rec{6}{The HEP Congressional advocacy group and APS DPF executive committee should identify an existing community group or create a new one to take ownership of strengthening connections between the HEP community and science and physics societies, including APS, AIP, and AAAS, and continuing conversations about how the HEP community can engage in advocacy for topics beyond HEP funding.}{The purpose of strengthening these connections should be to leverage external expertise on advocating for DEI, immigration policy, R\&D, basic science funding reform, and other areas that impact HEP. Maintaining strong connections between the HEP community and these external groups will expand the resources easily available to HEP community members in these areas. The HEP group tasked with strengthening these connections to external groups should also facilitate an expanded discussion around the potential for HEP-specific community advocacy in these areas.}
\end{recs}

\section{Community Interactions with the Funding Agencies}   
\label{sec:funding_agencies}

The inner workings of the funding agencies are highly relevant to HEP research and to individual HEP researchers, particularly as they progress through the earlier stages of career development. Robust channels of communication must exist between the funding agencies and the HEP community, and existing mechanisms are insufficient. The specifics of the granting processes at each of DOE and NSF, and expanding opportunities to improve feedback mechanisms regarding them, are of particular note within the community. Additionally, the power dynamic present between researchers and the funding agencies presents a substantial barrier to all interactions between them which needs to be addressed.

\begin{recs}
    \rec{7}{DOE and NSF should improve existing communication channels and create new ones, as necessary, to enable improved HEP community feedback to the funding agencies.}{HEP community members have substantive opinions about how our community interfaces with the funding agencies and lack sufficient avenues to provide feedback to them. Any steps taken to improve communication should include the development of mechanisms for community-level, individual, and anonymous feedback to be provided on topics including the climate of the field and the granting process. Transparency about what feedback is provided and what actions are taken to address feedback should be a key element of the solution.}
\end{recs}

The remainder of this section describes the current state of channels for communication between the HEP community and the funding agencies, followed by specific feedback about known deficiencies with recommendations for how to address them.

\subsection{Existing Mechanisms for Communication with the Funding Agencies}
\label{sec:funding_agency_feedback_mechanisms}

There exist groups, such as HEPAP and the Committees of Visitors (CoVs), that have the explicit mandate to liaise between the HEP community and the funding agencies. Although the membership of these groups is not community-driven (\textit{e.g.} through election), their compositions are specifically chosen to reflect the demographics of the field, and their members are generally well-respected within the HEP community. The primary role of these groups, as defined by the funding agencies, is to represent community interests to the funding agencies. While in principle their existence provides a conduit for individual community members to provide feedback to the funding agencies, in practice the available methods are flawed. HEPAP holds regular public meetings, at which time is always explicitly allocated for public comments. While this allows individual community members to speak directly to the funding agencies, the nature of these meetings (in particular, the attendance of Congressional and executive branch staff members) in practice limits the nature of feedback that individuals are comfortable sharing. Community members also may provide feedback directly to individual HEPAP or CoV members to be passed on to the funding agencies, however this avenue for feedback is biased toward later-career-stage individuals in the field that are more likely to have personal relationships with members of these groups.

Regular meetings between Principal Investigators (PIs) and program managers at the funding agencies are a vital feature of the relationship between the funding agencies and the HEP community and provide another mechanism for direct feedback to be provided by community members. At these meetings, the program managers describe current funding opportunities, changes in funding opportunities compared to previous years, and present the overall state of funding for HEP, and PIs are given an opportunity to directly provide feedback on the granting process or any other topics to the funding agencies. Participation in these meetings is generally restrictive enough that it excludes individuals (\textit{e.g.} early career community members applying for faculty positions) that may be interested in attending and could benefit. Simultaneously these meetings are well-attended enough to potentially disincentivize attendees from providing negative feedback because of how they may be perceived negatively by other community members, thereby adversely impacting their potential for career advancement. We note that many PIs additionally organize one-on-one meetings with their program managers before submitting grant applications and that program managers organize community meetings, at APS DPF for example. 

The merit and comparative grant review processes internal to the funding agencies also provide feedback mechanisms. DOE and NSF review panels provide explicit avenues for soliciting feedback from grant applicants and grant reviewers as part of their processes. However, we note that both applicants and reviewers may not feel comfortable giving negative feedback to the funding agencies because they may believe that it will negatively impact their current and future grant applications, respectively. Input received is shared with the Office of HEP, in the case of DOE grant reviews, and with the NSF Physics Division, in the case of NSF grant reviews. Larger-scale aspects of the application and review processes are managed at the agency or executive branch level, which may raise the bar for feedback to propagate up and for responses to propagate back to the HEP level. Additionally, DOE and NSF are distinct agencies with distinct granting processes, structures, and requirements. Feedback leading to change in one agency may not effect change in the other. For example, NSF explicitly solicits applicants to comment on their outreach activities, while the DOE does not. Finally, we note that the feedback mechanisms built into both agencies' review processes do not provide a means to supply feedback anonymously.

Snowmass also provides a mechanism for community feedback to be provided to the funding agencies, and the Community Engagement Frontier (CEF) in this current Snowmass cycle has taken advantage of this opportunity to generate critical feedback and recommendations. As discussed earlier in this report, there is reasonable concern that the impact of these recommendations will be minimized by the lack of an infrastructure designed to receive them. While the P5 panel is charged with distilling the physics recommendations of Snowmass into an actionable plan, there is no comparable entity in place to advance CEF recommendations. 

\subsection{Steps to Improve Communication with the Funding Agencies}
\label{sec:funding_agencies_comm_improvement}

Throughout the proceedings of CEF06, many community members expressed their feeling that there are limited mechanisms available for them to provide feedback to the funding agencies or that the existing tools are insufficient (as noted above). The deficiencies in available communication mechanisms all arise from a small number of underlying issues that should be addressed directly.

The power dynamics between researchers and representatives of the funding agencies, as well as, between researchers of different career stages poses a fundamental barrier to the free exchange of constructive criticism. It is important to recognize that these power dynamics are especially acute for researchers that are members of marginalized communities. A key element of a solution to this problem is to provide anonymous methods for feedback. Another element that should be considered is the utilization of existing community resources (\textit{e.g.} user groups, the DPF executive committee, and collaboration spokespeople) to facilitate communication between individual community members and the funding agencies. 

\begin{recs}
    \subrec{7.1}{DOE and NSF should work to reduce barriers created through the perceived power dynamic between community members and funding agency representatives. Simple communication paths should be created and widely advertised within the research community, including mechanisms that enable anonymous feedback.}
\end{recs}

It is important for early-career members of the community to fully participate in the long-term planning of our field, yet few dedicated channels exist to facilitate their communication with the funding agencies. Early-career researchers are less likely to be aware of existing communication channels, and in many cases the failings of the available channels (as noted above) pose even greater barriers to them. Many graduate students and postdocs would benefit from opportunities to interact directly with representatives of the funding agencies and to learn about how funding for HEP works. The postdoc-to-faculty transition is one example of a topic which is uniquely relevant to both early-career researchers and the funding agencies. We encourage the funding agencies to expand access to all existing communication channels to all members of the research community.

\begin{recs}
    \subrec{7.2}{DOE and NSF should take actions to facilitate communications with HEP researchers at all career stages, with an emphasis on expanding resources available to early-career researchers.}
\end{recs}

The specifics of the granting processes at DOE and NSF was a recurring topic through the proceedings of multiple topical groups in CEF. Many community members have opinions about how these processes could be improved but feel that there are insufficient channels available to them to provide feedback on this topic to the funding agencies. Additionally, there is broad confusion about which aspects of the granting process are mandated by the agencies or executive branch and which are legislated by Congress. Within the proceedings of CEF06, the granting process was discussed in a number of contexts, including DEI and accessibility, marginalized communities, and the emphasis placed on community engagement activities in grant applications. A detailed summary of these discussions can be found in Section 2.4 of Ref.~\cite{cef06paper3} and Section 7.2 of Ref.~\cite{cef06paper1} discusses empirical findings of bias in the grant review process. We recommend that the funding agencies facilitate dedicated conversations in this area with the HEP community, so that these opinions can be conveyed directly.

\begin{recs}
    \subrec{7.3}{DOE and NSF should facilitate dialogue with the HEP community about their granting processes and provide explicit opportunities for feedback in this area to be provided by community members.}
\end{recs}
\section{Expanding the Scope of HEP Government Engagement} 
\label{sec:expanded_engagement}

The proceedings of CEF06 provided a unique opportunity for the policy-minded subset of the HEP community to reflect on how we, as a community, engage with the government at all levels. Critically, these conversations occurred outside the scope of a particular year's advocacy activities, which provided the space for longer-term, bigger-picture, and farther-reaching ideas to be discussed. A high-level summary of these discussions is presented below along with targeted suggestions for how to continue exploring these ideas. A more complete summary of the discussion about these topics and others is available in Sections 3-5 of Ref.~\cite{cef06paper3} and Section 7 of Ref.~\cite{cef06paper1}.

\subsection{Executive Branch Engagement}

As part of the annual community-driven advocacy activities, community members meet with two agencies within the Executive Office of the President (EOP): the Office of Science and Technology Policy (OSTP) and the Office of Management and Budget (OMB). OSTP's role is to advise the President and others within EOP on science and technology policy matters. OMB's role is to allocate funding to executive branch agencies in line with Congress's authorizations and appropriations bills. OSTP and OMB release an annual joint memorandum of their budget priorities -- this document serves as a driver of the conversation between representatives of the HEP community in the meetings with these agencies. In these meetings the details and justification for the HEP appropriations requests are discussed, and community members provide a summary of our field's status and top priorities for the coming year. The staffing within these agencies is often tied to a specific Administration, so there is generally high turnover. Thus, meeting regularly with the agencies provides a good opportunity to inform new staff about HEP priorities. The meetings also provide an opportunity to learn about the current Administration's priorities and to discuss the content of the preceding year's joint memorandum.

While these meetings in principle cover the same topics as the meetings between community members and Congressional staff, there are important differences reflective of the distinct priorities of the executive and legislative branches of the federal government. At present, our strategy is not optimized to fully account for these differences, though there are well-defined steps that could be taken to make this optimization and thereby increase the impact of this effort.

The annual advocacy trip to Washington, D.C. is timed to have maximal impact on the Congressional stage of the budget cycle, meaning that these meetings with OSTP and OMB occur \textit{after} the time that would have the maximal impact on the executive stage of the budget cycle (\textit{i.e.} when the President's Budget Request is being developed). Given sufficient resources, or perhaps by transitioning to a virtual model for these executive branch meetings, our community could meet with both legislative and executive branches at the optimal times for each.

The materials used in these meetings are currently the standard HEP communication materials described in Section \ref{sec:outreach_materials}, which are targeted primarily at a general audience, rather than at policy experts. Additionally, the development of a strategy for these meetings is generally done within a subset of the ``DC trip'' organizers without additional input from government relations experts beyond what has been collected already for the Congressional advocacy effort. The production of new outreach materials and messaging points targeted specifically at the executive branch agencies, along with specific training on the role of these groups provided to meeting attendees, would significantly benefit our engagement in this area. 

The impact of community interactions with executive branch agencies can be high. Prior to the development and successful implementation of the 2014 P5 report, it was feedback from these agencies about the importance of unity in our field that led the HEP community to rally around the 2014 P5 plan. The 2013 Snowmass process, the 2014 P5 report, and the robust community unity around its implementation were direct responses to this feedback, and have been perceived positively. While it is difficult to isolate and quantify the impact of the community response to this feedback, we do know that OHEP funding in the President's Budget Request increased between 2015 and 2017 (see Fig. \ref{fig:OHEP_fun}). 

\begin{recs}
\rec{8}{The HEP Congressional advocacy group should work to improve HEP community engagement of the executive branch, especially OMB and OSTP.}{Steps should be taken to optimize community-driven advocacy targeted at these groups and to separate the logistics of this effort from the community-driven advocacy targeted at Congress. Meetings with these agencies should be timed for maximum impact on the development of the President's Budget Request. Specialized outreach materials and messaging points should be developed, and HEP community members meeting with these agencies should receive specialized training.}
\end{recs}

\subsection{State and Local Government Engagement}

There is currently no HEP community-driven advocacy or systematic engagement of government bodies in the U.S. below the federal level, \textit{i.e.} at the state and local levels. Yet, the relationships between HEP institutions and facilities and their state and local governments are important. These institutions and facilities in many cases do have formal relationships with their government representatives at all levels, ranging from governors to aldermen. An example of a group with the express purpose of facilitating this type of relationship is the Fermilab Community Advisory Board, that ``provides ongoing advice and guidance related to the future of the laboratory. The Board gives feedback on proposed new projects, reviews planned construction activities, advises Fermilab on all forms of public participation, and acts as a liaison with local organizations and communities.'' \cite{FCAB}. Strong relationships between HEP facilities and their local communities are highly beneficial to both the HEP community and to the general public -- this is expanded on in Section 3 of Ref.~\cite{cef07} and in Ref.~\cite{Zens:2022lnv}.

During the proceedings of CEF06, the potential for expanding HEP community-wide advocacy to the state and local levels was discussed. There are many cases where existing tools and resources could be adapted to such new engagement with minimal effort. There is precedent from localized examples which indicates that increased engagement between the HEP community and state and local government officials has been beneficial. We recommend that steps be taken to develop a standard HEP community approach to engaging state and local governments. An effort should be undertaken to understand the potential direct impacts and the resources needed for the organized engagement of state and local governments relevant for key HEP facilities, including Fermilab, SLAC, BNL, and SURF.

\begin{recs}
    \rec{9}{The HEP Congressional advocacy group and APS DPF executive committee should facilitate discussions to explore the potential advantages to systematic engagement of local and state governments.}{The goal of these discussions should be understanding the potential costs and benefits of allocating resources for such activities. Specific locations of particular relevance to HEP activities should be identified to guide these discussions.}
\end{recs}

\subsection{Rethinking Basic Research in the U.S.}

The majority of conversations that occurred throughout the proceedings of CEF06 were focused on the potential for directly expanding existing HEP community government engagement activities to increase their collective depth (\textit{e.g.} investing additional resources in the annual D.C. trip) and breadth (\textit{e.g.} exploring options for improved communication with the funding agencies). However some time was also spent discussing the role that the HEP community could play in broader movements within the U.S. scientific community related to fundamental aspects of how research is funded and prioritized in this country. While we do not assert any specific recommendation in this area, we feel that the ideas themselves merit broader dissemination and hope that they can be robustly discussed within the community in the future. Two examples of such topics are discussed below.

First, federal funding for scientific research in the U.S., despite growing in absolute terms, has not kept up with economic growth or with international trends. This is reflected in funding for research and development (R\&D) in the U.S., which has been decreasing as a fraction of GDP for over thirty years, while investments into research have drastically increased in foreign nations such as China~\cite{cef06paper1}. These facts, along with the known substantial economic returns from investments in science funding~\cite{SALTER2001509,summers}, merit consideration of substantially expanding the funding for scientific research beyond present levels. As noted earlier (see Section \ref{sec:implicit_advocacy}), HEP community-driven advocacy already includes implicit support of basic research as a national priority, though present activities fall far short of the dramatic increase in investment in this area that is advocated by others.

Second, the processes used by funding agencies to award grants to researchers are known to be imperfect.  HEP community members are aware that the bureaucracy involved with applying for and managing grants consumes substantial resources -- this is a trend that has been studied and well characterized within the broader scientific community~\cite{LINK2008363,Bozeman2015,workload2014reducing,fdp2018}. Studies of granting processes have additionally shown significant biases inherent in the peer-review systems used to review grant applications~\cite{Cole1981,Graves2011,Fang2016,Pier2018,Ginther2011,Tabak2011,Witteman2019,impact2007,Daniels2015,DOE2019,Lauer2017,Conix2021}. Proposals of alternate schemes that could be implemented to award grants exist, including equally dividing all available funds amongst the pool of all eligible researchers~\cite{Vaesen2017} and lotteries to randomly award grants amongst eligible applicants~\cite{Avin2015,Fang2016lottery,Adam2019}. HEP community-driven advocacy presently do not include any conversations with Congressional offices about the inner workings of the funding agencies, beyond the highest levels of how funding is distributed to support HEP research.

Both of these examples are topics which should be widely and openly discussed within the HEP and U.S. research communities. Advocating for such big picture changes in how science works in the U.S. is an area where the HEP community could assume a leading role in the broader scientific community by leveraging our significant expertise in government engagement and in particular in interacting with Congress. These topics and others are discussed in greater detail in Section 7 of Ref.~\cite{cef06paper1}.
\section{Conclusion}   
\label{sec:conclusion}

We have presented a summary of the findings and recommendations of the Snowmass group focused on Public Policy \& Government Engagement, Community Engagement Frontier topical group \#6 (CEF06). The charge of CEF06 was to review how the HEP community engages with government at all levels and all related topics. Among the topics investigated were how public policy impacts members of the HEP community and the HEP community at large and the issue of awareness within the HEP community of direct community-driven engagement of the U.S. federal government. The proceedings of CEF06 through 2020 and 20221 culminated in three Snowmass contributed papers \cite{cef06paper1,cef06paper2,cef06paper3}. The contents of these papers are summarized in this report and distilled into 9 major recommendations and 18 additional subrecommendations.

The authors of this report, the conveners of CEF06, would like to thank all of the HEP community members who participated in the proceedings of our group, and to all community members that participated in the efforts of the Community Engagement Frontier overall. Work in this area is woefully undervalued, but critical for the survival of our field. We would specifically like to thank the conveners of the Community Engagement Frontier who have dedicated themselves for the past three years to this important work and who have championed the value of community engagement activities throughout the entire Snowmass process.
\clearpage

\bibliographystyle{JHEP}
\bibliography{Bibliography/common,Bibliography/main}

\end{document}